# Crystal structure and charge order below the metal-insulator transition in the vanadium bronze β-SrV₆O₁₅

Claire Sellier, Florent Boucher & Etienne Janod

*Institut des Matériaux Jean Rouxel, UMR 6502 CNRS-Université de Nantes, 2 rue de la Houssinière, BP 32229, 44322 Nantes Cedex 3, France*

**Abstract**

Single crystal X-ray diffraction measurements were performed on one-dimensional mixed valence vanadium bronze β-SrV₆O₁₅ in which a metal-insulator transition exists at 170 K. Above 170 K the P2₁/a structure with zigzag order of Sr in the tunnels of vanadium is confirmed. In the structure below 170 K, the P2₁/a space group is retained with a *b*-axis threefold increase and a charge order appears. A Bond Valence Sum analysis shows that the charge order may consist in clusters of $V^{4+}$ regularly spaced along 1D direction.

## 1. Introduction

The vanadium bronzes β-$A_x$V₆O₁₅ ($A$ = Li⁺, Na⁺, Ag⁺, Cu⁺, Pb²⁺, Ca²⁺, Sr²⁺; 0.5 ≤ x ≤ 1) have been discovered more than thirty years ago [1,2] and have attracted much attention in recent years. This interest is related to the unusual physical properties of these mixed valence $V^{4+}$ (3d¹) / $V^{5+}$ (3d⁰) compounds, such as their quasi one-dimensional (1D) conductor character at high temperature. For the stoichiometric composition x = 1, most of the above compounds undergo a metal-insulator transition (MIT) associated with a charge order at temperature ranging from $T_{MIT}$ = 90 K (AgV₆O₁₅ [3]) to 180 K (LiV₆O₁₅ [4]). The high temperature metallic phase is stabilized under pressure and, surprisingly, superconductivity up to 8 K was recently discovered in NaV₆O15 at high pressure (8 GPa) [5,6]. At ambient pressure, the magnetic properties in the charge ordered state depend on the valence of $A$ cation. For most of the compounds with a monovalent $A$ cation ($A$ = Li⁺, Na⁺, Ag⁺), the behavior is Curie-Weiss-like below $T_{MIT}$ and an antiferromagnetic long-range order (AFLRO) is established at lower temperature [7]. The magnetism of the compounds with a divalent $A$ cation is more puzzling, especially for CaV₆O₁₅ ($T_{MIT}$ = 150 K) and SrV₆O₁₅ ($T_{MIT}$ = 170 K) for which there is no sign of long range magnetic order down to 2 K. A singlet ground state with a gap in the spin excitation spectrum even appears for SrV₆O₁₅ [8]. The difference in the behavior of compounds with monovalent and divalent $A$ cation has not been clarified yet but it is probably related with the charge ordered pattern appearing below the MIT. The latter has however not been determined yet in SrV₆O₁₅. In this context, the determination of the structure of SrV₆O₁₅ above and below the metal-insulator transition are clearly of large interest, since it could provide valuable indications on the charge ordering process through a Bond Valence Sum [9] analysis.

The room temperature (RT) structure of the compounds β-$A_x$V₆O₁₅ was initially determined by Wadsley [1] for $A$ = Na and x = 1. It crystallizes in a monoclinic cell in the space group C2/m (*a* = 10.088(3) Å, *b* = 3.617(2) Å, *c* =15.449(3) Å, *β* = 109.57(2) °). Bouloux et al [2] confirmed latter that SrV₆O₁₅ is isostructural with NaV₆O₁₅. The structural bricks are strongly distorted VO₆ octahedra, with a short vanadyl V=O bond (≈ 1.60 Å), average of four V-O distances (1.85 - 2 Å) in the plane perpendicular to the vanadyl bond and a long V-O bond (> 2.20 Å) opposed to the vanadyl bond. When this last distance is longer than 2.5 Å, a VO₅ pyramid description is preferred. Three independent sites of vanadium exist in this structure, as shown in Fig. 1. In the *ab* plane, V1 and V2 sites form a layer of VO₆ octahedra, in which the V2 sublattice forms two-leg ladders and the V1 sublattice forms zig-zag chains, both running along the *b* axis. The third vanadium V3 form zigzag chains of VO₅ pyramids which bridge two adjacent layers together and delimit tunnels parallel to the *b* direction. In the C2/m space group, the Sr atoms are randomly distributed over two available sites in the tunnels with an occupancy x/2. The two Sr sites are 1.98 Å apart in the same *ac* plane. So, for steric reasons, they can not be occupied at the same time by Sr²⁺ ions.

However, it turns out that satellite reflections at q = 1/2 *b** always appear for stoichiometric (x=1) or nearly stoichiometric compounds [10,11]. For this reason, the average description with the C2/m space group is only acceptable for off-stoichiometric compounds (x < 0.9). Recently, the real ordered structures of NaV₆O₁₅ and SrV₆O₁₅ have been refined above their MIT [10]. These compounds crystallize with the same structural type (space group P2₁/a) with a unit cell along the *b* direction twice as large compared to the C2/m one. This high temperature structure will be hereafter referred to P2₁/a (2*b*). The origin of the cell doubling is the zigzag ordering of the Na⁺ (Sr²⁺) ions in the tunnels, now located on fully occupied sites. Also, it was recently observed that a structural phase transition was associated with the MIT in NaV₆O₁₅ [10]. The low temperature structure retains the same space group (P2₁/a), but with a *b* cell parameter three times larger than above the transition. The low temperature structure of SrV₆O₁₅ has however not been determined yet.

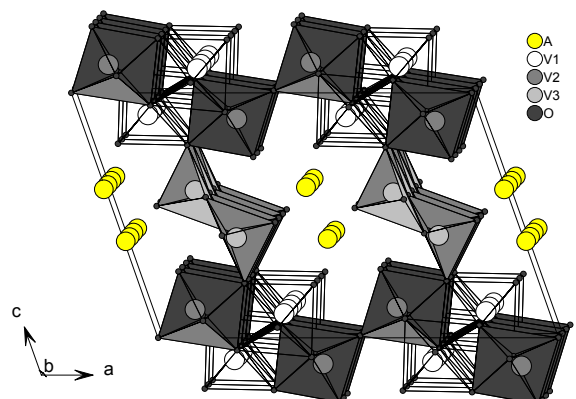

Figure 1: View of the average structure of $A_x$V₆O₁₅ in the C2/m space group. The occupation ratio of the $A$ site is x/2.

In this paper, we first confirm the high temperature structure of SrV₆O₁₅ as described in reference [10]. Then, we present the first structural determination of the charge ordered phase of SrV₆O₁₅ at low temperature. Moreover, a Bond Valence Sum analysis (BVS) is performed on this low temperature (LT) structure and a charge ordered pattern very different from the one recently proposed for NaV₆O₁₅ [10] is given.

## 2. Experimental part

### Synthesis

A starting powder was synthesized by mixing ad hoc amounts of SrV₂O₆, V₂O₅ and V₂O₃ in an evacuated and sealed silica tube and by heating the latter at 650°C during 33 hours. SrV₂O₆ was prepared by a solid state reaction of SrCO₃ (99.995 % / Aldrich) with V₂O₅ (99.6 % / Aldrich) in air at 600°C, and V₂O₃ was prepared by reducing V₂O₅ under flowing hydrogen at 900°C. X-Ray powder diffraction experiments indicate that the obtained powder is a mixture of SrV₆O₁₅ and Sr₂V₃O₉. Single crystals were grown by a self-flux method in a furnace with a controlled partial pressure of oxygen. Semi-quantitative Energy Dispersive X-ray Spectroscopy (EDXS) analyses of some single crystals by means of a Jeol microscope (PGT-IMIX-PTS equipped Jeol-JSM5800LV) gave the approximate Sr : V = 0.98 ± 0.05 : 6.00 ratio.

### Data collection and data reduction

For data collection, a needle shape single crystal of SrV₆O₁₅ with dimensions 0.4×0.05×0.01 mm was glue with nail polish at the tip of a lindemann glass capillary. Reflection intensities were collected on a Nonius Kappa CCD [12] diffractometer equipped with a two-dimensional CCD detector (diameter 9 cm), a conventional X-ray tube (Mo K-L₂,₃ radiation, λ = 0.71073 Å) and an Oxford Cryostream low-temperature device. The RT data collection have been done in about 7 hours in a rather standard way: detector distance 25 mm, $\Theta_{max}$ = 40°, 304 frame, φ(or ω) scan, 1.7° rotation per frame, 76s per frame. For the LT measurement (90K), a different strategy was used in order to get a good I/σ(I) ratio for the small satellite reflections. The whole data collection took nearly 3 days: detector distance 38 mm, $\Theta_{max}$ = 37.5°, 544 frames with 1.0° φ(or ω) rotation and 450s per frame. A fast scan was used to solve the problem of overflow for the strong reflections: 187 frames with 1.0° φ(or ω) rotation and 45s per frame. The intensity of each reflection was integrated using the Eval CCD package [13] and corrected for Lorentz and polarization effects. All supplementary treatments (numerical absorption corrections and averaging process) and least square refinements were carried out with the JANA2000 software package [14]. For the absorption corrections, an optimization of the crystal shape was done with the X-Shape program [15] using equivalent reflections. Details on data collections and processes are given Table I.

### Structure refinement

**RT phase**: As previously shown by Yamaura et al. [10], the diffraction pattern of the RT phase of SrV₆O₁₅ can be indexed using the space group P2₁/a and the following unit cell: a ≈ 15.44 Å, b ≈ 7.30 Å, c ≈ 10.16 Å, β ≈ 109.4. The starting atomic positions were derived from the average structure C2/m of Wadsley [1], considering an ordering on the Sr sites. The refinement was based on F² and a secondary isotropic extinction coefficient was refined. With all the atoms refined anisotropically, the final R(obs)/R_w(obs) = 3.08/6.35 was obtained. No significant residue was found in the Fourier difference map. The refined atomic positions and equivalent atomic displacement parameters (ADP) for the RT phase are given in Table II. The occupation factor of the Sr has been refined and the shift from x = 1 was found to be smaller than 0.01, in good accordance with the EDXS result.

**LT phase**: Looking carefully at the extinction rules, it was found that the diffraction pattern of the LT phase can be indexed with the same space group P2₁/a considering a unit cell three times larger: a ≈ 15.41 Å, b ≈ 21.88 Å, c ≈ 10.15 Å, β ≈ 109.4°. The atomic positions were derived from RT unit cell considering a supercell: $b_{LT}$ = $3b_{RT}$. Solving first the average structure, it was observed that small in-plane displacements on the V2 sites are mainly responsible for the satellite reflections at LT (see below the structure description). So, the starting solution for the refinement was to introduce off centered V2 sites. Considering the six different possibilities for the displacements, a good explanation of the satellite intensities was found for only one set. All the atoms were refined anisotropically but constraints were applied on atomic displacement parameters (ADP) of oxygen sites in order to keep the average symmetry of the RT phase. Moreover, a damping factor of 0.5 was used to prevent the divergence of the refinements. With 7091 observed reflections (3138 observed satellites) and 416 refined parameters, the following reliability factors were obtained: R(obs)/R_w(obs) = 3.61/6.67 with for the main reflections: R(obs)/R_w(obs) = 1.95/5.32 and for the satellites: R(obs)/R_w(obs) = 11.41/12.82. The refined atomic positions and equivalent ADP for the LT phase are given in Table III.

Table I: Data, Intensity Measurement, and Structure Refinement Parameters for $SrV_6O_{15}$

| | T = 90 K | T = 300 K |
|---|---|---|
| Physical, crystallographic, and analytical data | | |
| Formula | $SrV_6O_{15}$ | $SrV_6O_{15}$ |
| Crystal system | monoclinic | monoclinic |
| Space group | $P2_1/a$ (no. 14) | $P2_1/a$ (no. 14) |
| Cell parameters | | |
| a (Å) | 15.4110(5) | 15.4356(15) |
| b (Å) | 21.8809(12) | 7.3025(7) |
| c (Å) | 10.1500(9) | 10.1632(11) |
| β (°) | 109.42(8) | 109.41(10) |
| V (Å³) | 3227.98(4) | 1080.46(2) |
| Z | 12 | 4 |
| Density calc. (g cm⁻³) | 3.909 | 3.893 |
| Data collection | | |
| Diffractometer | Nonius Kappa CCD | Nonius Kappa CCD |
| Radiation | Mo $K$ L$_{2-3}$ (0.71073 Å) | Mo $K$ L$_{2-3}$ (0.71073 Å) |
| μ (mm⁻¹) | 0.9988 | 0.9998 |
| Crystal size (mm) | 0.4×0.05×0.01 | 0.4×0.05×0.01 |
| Total reflections | 70469 | 29184 |
| Observed reflections I > 2σ(I) | 7091 | 3574 |
| Data reduction | | |
| Independent reflection | 16757 | 5553 |
| $R_{int} = \Sigma|I - I_{avl}|/\Sigma I$ | 0.0495 | 0.0602 |
| Absorption correction. | Gaussian integration method | Gaussian integration method |
| Transmission coeff. | 0.244-0.874 | 0.234-0.893 |
| Independent reflections with I>2.0 σ($I$) | 7091 | 3574 |
| Refinement[a] | | |
| Refinement | $F^2$ | $F^2$ |
| R (%) | 3.61 (all), 1.95 (main), 11.41 (satellite) | 3.08 |
| $R_w$ (%) | 6.67 (all), 5.32 (main), 12.82 (satellite) | 6.35 |
| No. of refined parameters | 416 | 200 |
| Difference Fourier residues (e⁻/Å³) | -2.02, +2.42 | -1.07, +0.99 |
| Goodness of fit | 1.59 | 1.31 |
| Secondary extinction | 1.70(6) | 2.11(7) |

[a]$R = \Sigma\ [|F_o - F_c|]/\Sigma\ [|F_o|]$ and $R_w = [\Sigma w\ (|F_o| - |F_c|^2)/\Sigma w F_o^2]^{1/2}$ with w = 1/σ($F_o$)

Table II: $SrV_6O_{15}$ T = 300 K: atomic positions and equivalent atomic displacement parameters

| Atom | x | y | z | U$_{eq}$ (10⁻² Å²) |
|---|---|---|---|---|
| Sr | 0.253999(14) | 0.62439(3) | 0.40470(2) | 1.181(6) |
| V1a | 0.58640(2) | 0.62522(11) | 0.10116(3) | 0.641(9) |
| V1b | 0.58478(2) | 0.12413(11) | 0.09850(3) | 0.634(8) |
| V2a | 0.36576(2) | 0.62383(11) | 0.11132(3) | 0.751(9) |
| V2b | 0.37050(3) | 0.12558(11) | 0.11982(3) | 0.975(10) |
| V3a | 0.53323(2) | 0.62789(11) | 0.40747(3) | 0.650(9) |
| V3b | 0.53769(2) | 0.12149(11) | 0.41223(3) | 0.667(8) |
| O1a | 0.25082(9) | 0.6267(6) | 0.00464(14) | 1.21(4) |
| O2a | 0.06235(10) | 0.6254(4) | 0.05290(15) | 0.95(4) |
| O2b | 0.06674(10) | 0.1263(4) | 0.04874(14) | 0.87(4) |
| O3a | 0.88379(10) | 0.6226(4) | 0.07739(14) | 0.77(4) |
| O3b | 0.88372(10) | 0.1279(4) | 0.07543(14) | 0.76(4) |
| O4a | 0.68714(11) | 0.6324(4) | 0.21862(16) | 1.29(4) |
| O4b | 0.68518(11) | 0.1161(4) | 0.21697(16) | 1.24(4) |
| O5a | 0.51575(10) | 0.6211(5) | 0.22115(14) | 0.78(4) |
| O5b | 0.51353(10) | 0.1258(5) | 0.22042(13) | 0.80(4) |
| O6a | 0.36386(11) | 0.6170(5) | 0.27266(16) | 1.14(4) |
| O6b | 0.35797(11) | 0.1252(5) | 0.26986(15) | 1.44(5) |
| O7a | 0.00041(10) | 0.6210(4) | 0.41893(14) | 0.81(4) |
| O7b | 0.01424(11) | 0.1239(4) | 0.43112(15) | 1.06(4) |
| O8a | 0.64559(11) | 0.6420(4) | 0.47153(16) | 1.17(5) |
| O8b | 0.64994(11) | 0.1070(4) | 0.47037(16) | 1.15(5) |

Table III: SrV$_6$O$_{15}$ T = 90 K: atomic positions and equivalent atomic displacement parameters

| Atom | x | y | z | U$_{eq}$ (10$^{-2}$ Å$^2$) |
|---|---|---|---|---|
| Sr1 | 0.25493(3) | 0.874866(19) | 0.40556(5) | 0.391(11) |
| Sr2 | 0.253373(17) | 0.54139(3) | 0.40581(3) | 0.385(7) |
| Sr3 | 0.25352(4) | 0.208453(15) | 0.40331(5) | 0.392(13) |
| V1a1 | 0.58769(7) | 0.87600(5) | 0.09811(11) | 0.25(2) |
| V1a2 | 0.58569(5) | 0.54316(6) | 0.10162(7) | 0.228(16) |
| V1a3 | 0.58574(6) | 0.20806(5) | 0.10311(11) | 0.23(2) |
| V1b1 | 0.58398(6) | 0.70843(5) | 0.10071(10) | 0.23(2) |
| V1b2 | 0.58311(7) | 0.37467(5) | 0.09999(11) | 0.21(2) |
| V1b3 | 0.58633(5) | 0.04100(6) | 0.09390(7) | 0.225(15) |
| V2a1 | 0.36200(6) | 0.87474(5) | 0.10850(11) | 0.20(2) |
| V2a2 | 0.37332(4) | 0.54154(7) | 0.11781(7) | 0.242(14) |
| V2a3 | 0.36257(6) | 0.20790(5) | 0.10777(10) | 0.24(2) |
| V2b1 | 0.37744(6) | 0.70878(5) | 0.12296(10) | 0.28(2) |
| V2b2 | 0.37686(7) | 0.37530(5) | 0.12304(11) | 0.24(2) |
| V2b3 | 0.36035(4) | 0.04201(7) | 0.11808(7) | 0.289(13) |
| V3a1 | 0.53276(7) | 0.87653(5) | 0.40365(11) | 0.19(2) |
| V3a2 | 0.53420(5) | 0.54299(6) | 0.41014(7) | 0.225(15) |
| V3a3 | 0.53227(6) | 0.20897(5) | 0.40742(10) | 0.26(2) |
| V3b1 | 0.53980(6) | 0.70777(5) | 0.41779(10) | 0.26(2) |
| V3b2 | 0.53977(7) | 0.37346(5) | 0.41768(11) | 0.24(2) |
| V3b3 | 0.53532(5) | 0.04063(6) | 0.40442(7) | 0.236(15) |
| O1a1 | 0.2506(2) | 0.8744(2) | 0.0109(4) | 0.50(2) |
| O1a2 | 0.25176(12) | 0.5422(3) | -0.0068(2) | 0.50 |
| O1a3 | 0.2503(3) | 0.2093(3) | 0.0078(5) | 0.50 |
| O2a1 | 0.0610(3) | 0.8741(2) | 0.0535(4) | 0.37(2) |
| O2a2 | 0.06196(17) | 0.5414(3) | 0.0559(3) | 0.37 |
| O2a3 | 0.0625(3) | 0.2085(2) | 0.0539(4) | 0.37 |
| O2b1 | 0.0681(3) | 0.7087(2) | 0.0477(4) | 0.36(2) |
| O2b2 | 0.0667(3) | 0.37561(19) | 0.0460(4) | 0.36 |
| O2b3 | 0.06469(17) | 0.0431(3) | 0.0463(3) | 0.36 |
| O3a1 | 0.8823(3) | 0.8726(2) | 0.0778(5) | 0.38(2) |
| O3a2 | 0.88459(19) | 0.5418(3) | 0.0794(3) | 0.38 |
| O3a3 | 0.8844(3) | 0.2079(2) | 0.0769(4) | 0.38 |
| O3b1 | 0.8843(3) | 0.7101(2) | 0.0752(4) | 0.31(2) |
| O3b2 | 0.8851(3) | 0.37636(19) | 0.0778(5) | 0.31 |
| O3b3 | 0.8816(2) | 0.0446(2) | 0.0748(3) | 0.31 |
| O4a1 | 0.6885(3) | 0.87949(19) | 0.2164(4) | 0.48(3) |
| O4a2 | 0.68597(19) | 0.5451(2) | 0.2214(3) | 0.48 |
| O4a3 | 0.6878(2) | 0.2108(2) | 0.2189(4) | 0.48 |
| O4b1 | 0.6856(3) | 0.7064(2) | 0.2181(5) | 0.57(2) |
| O4b2 | 0.6843(3) | 0.3733(2) | 0.2166(5) | 0.57 |
| O4b3 | 0.68590(19) | 0.0387(3) | 0.2150(3) | 0.57 |
| O5a1 | 0.5169(3) | 0.8748(2) | 0.2240(5) | 0.38(2) |
| O5a2 | 0.51405(19) | 0.5409(3) | 0.2229(3) | 0.38 |
| O5a3 | 0.5168(3) | 0.2075(2) | 0.2201(4) | 0.38 |
| O5b1 | 0.5137(3) | 0.7105(2) | 0.2193(4) | 0.33(2) |
| O5b2 | 0.5132(3) | 0.37602(19) | 0.2188(5) | 0.33 |
| O5b3 | 0.51356(19) | 0.0436(2) | 0.2222(3) | 0.33 |
| O6a1 | 0.3640(3) | 0.8738(2) | 0.2727(5) | 0.54(2) |
| O6a2 | 0.36191(18) | 0.5408(2) | 0.2732(3) | 0.54 |
| O6a3 | 0.3642(3) | 0.2067(2) | 0.2720(4) | 0.54 |
| O6b1 | 0.3589(3) | 0.7105(2) | 0.2692(4) | 0.55(2) |
| O6b2 | 0.3579(3) | 0.3771(2) | 0.2679(5) | 0.55 |
| O6b3 | 0.35566(17) | 0.0432(3) | 0.2749(3) | 0.55 |
| O7a1 | -0.0006(2) | 0.87520(18) | 0.4168(4) | 0.36(2) |
| O7a2 | 0.00055(14) | 0.5408(2) | 0.4194(2) | 0.36 |
| O7a3 | -0.0011(3) | 0.20911(16) | 0.4163(4) | 0.36 |
| O7b1 | 0.0150(3) | 0.70714(16) | 0.4320(4) | 0.41(2) |
| O7b2 | 0.0138(2) | 0.37805(19) | 0.4320(4) | 0.41 |
| O7b3 | 0.01436(14) | 0.0420(2) | 0.4318(2) | 0.41 |
| O8a1 | 0.6451(2) | 0.88166(16) | 0.4704(4) | 0.47(3) |
| O8a2 | 0.64699(18) | 0.54812(18) | 0.4728(3) | 0.47 |
| O8a3 | 0.6449(2) | 0.21366(16) | 0.4717(4) | 0.47 |
| O8b1 | 0.6525(2) | 0.70255(16) | 0.4722(4) | 0.50(3) |
| O8b2 | 0.6523(2) | 0.36869(17) | 0.4731(4) | 0.50 |
| O8b3 | 0.64790(18) | 0.03602(19) | 0.4661(3) | 0.50 |

## 3. Results

### RT phase

Our structural refinement at RT is in accordance with the structure recently published [10]. The origin of the *b*-axis doubling is the zigzag ordering of the Sr which populate alternately the high and low Sr sites of the C2/m structure in the tunnels (see Fig. 2).

This intra-tunnel ordering can be directly detected from a difference Patterson (DP) map (or a Patterson map using only supercell reflections), as demonstrated recently for β-PbV$_6$O$_{15}$ [11]. However, we note that the absence of intensity in the DP map at vectors (1/2, 1/4, 0) and (1/2, 3/4, 0.195) is fully consistent with the P2$_1$/a space group and a perfect inter-tunnel ordering, contrary to what is claimed in ref [11]. This inter-tunnel order is already imposed by the C centering of the C2/m space group. Another striking consequence of the Sr zigzag order is the splitting of each vanadium site of the C2/m space group (V1, V2 and V3) into two different sites in the P2$_1$/a space group (V1a, V1b; V2a, V2b; V3a, V3b). In the average C2/m structure, the average number of Sr atoms in the first coordination sphere of each oxygen atom involved in a V=O vanadyl is: one half, one and two for oxygen atom linked with V2, V1 and V3 respectively. In the P2$_1$/a space group, due to the Sr ordering, these values become zero (V2b), one (V1a, V1b and V2a) or two (V3a and V3b). As a consequence, the two V1-sites (V1a and V1b) do not differentiate much from each other in their first coordination sphere and the same thing holds for the two V3-sites. On the other hand, the VO$_6$ polyhedra of the two V2-sites become very different (see Fig.3).

It is therefore expected that the most affected vanadium will be V2, which is confirmed in the V-O distances shown in Fig.3. More precisely, the coordination polyhedra of the vanadium V2a contains two short V-O distances (1.650 Å and 1.745 Å) and becomes close to the Trigonal Bipyramid (TB) geometry [16]. On the other hand, V2b remains much less distorted with only one vanadyl bond (see Fig. 3). Another striking consequence of the Sr ordering is a dimerization of the V3 vanadium along the *b*-axis (see Fig. 4). The modulation of the V3-V3 distances (≈ 0.1 Å) originates from an alternation of opened (≈ 153°) and closed (≈ 144°) V3-O-V3 angles without any significant change in the V3-O distances.

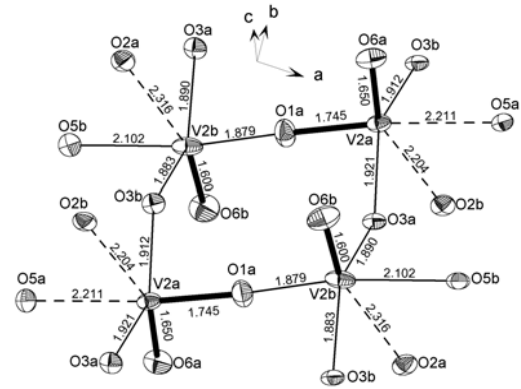

Figure 3: V-O distances around the V2a and V2b sites of the RT structure of SrV$_6$O$_{15}$. Short bonds (<1.77 Å) are indicated by thick lines and longer bonds by dash lines.

This dimerization effect is strongly attenuated in the V1 and V2 chains, with a V-V distances modulation smaller than 0.025 Å. The Bond Valence Sum (BVS) analysis [9] of the room temperature structure of SrV$_6$O$_{15}$ does not reveal any significant differences in the formal valences of the six different vanadium sites (see Table IV).

### LT phase

Between room temperature and 90 K, a structural phase transition occurs. Satellite reflections appear below the transition (170K) indicating a *b*-axis threefold increase compared to the P2$_1$/a (2*b*) room temperature cell (Fig. 5a). The structural refinement shows that the P2$_1$/a space group is retained. Therefore the low temperature phase will be referred to as P2$_1$/a (6*b*). In the later structure, three vanadium (labeled V1a1, V1a2, V1a3 and so on) issue from each vanadium of the room temperature P2$_1$/a structure (V1a, V1b, …), generating eighteen inequivalent V positions. In the V2 ladder, half of the V2 are close to the Trigonal Bipyramid (TB) environment. Groups of three consecutive V2 (TB) alternate with three less distorted V2 along the ladder legs (*b*-direction). The "second" short vanadyl V=O bond of each V2 (TB) is directed toward the oxygen shared by two V2 of the same rung (see Fig. 5b - 5c), in such a way that each group of three V2 (TB) is displaced toward the ladder axis. In comparison with the V2-sites, V1 and V3-sites are less modified. Only two V3 (V3a1 and V3b3, see Fig. 6) tend to distort slightly into the TB geometry. It is rather clear that the P2₁/a (6*b*) low temperature structure is closely related to the high temperature P2₁/a (2*b*). This is particularly true for the dimerized V3 chains which are essentially unaffected by the MIT (see Fig. 4).

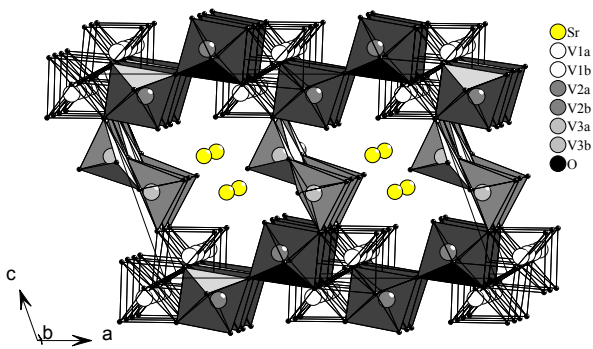

Figure 2: View of the ordered structure of SrV$_6$O$_{15}$ in space group P2$_1$/a at T = 300K.

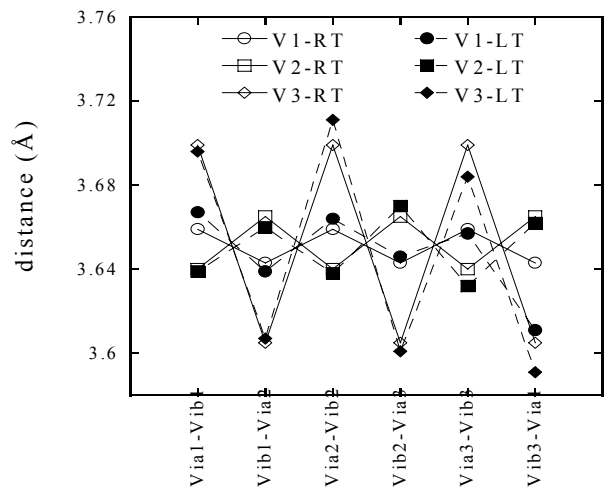

Figure 4: V-V distances along the *b*-axis at RT and LT (*i* corresponds to the V site number).

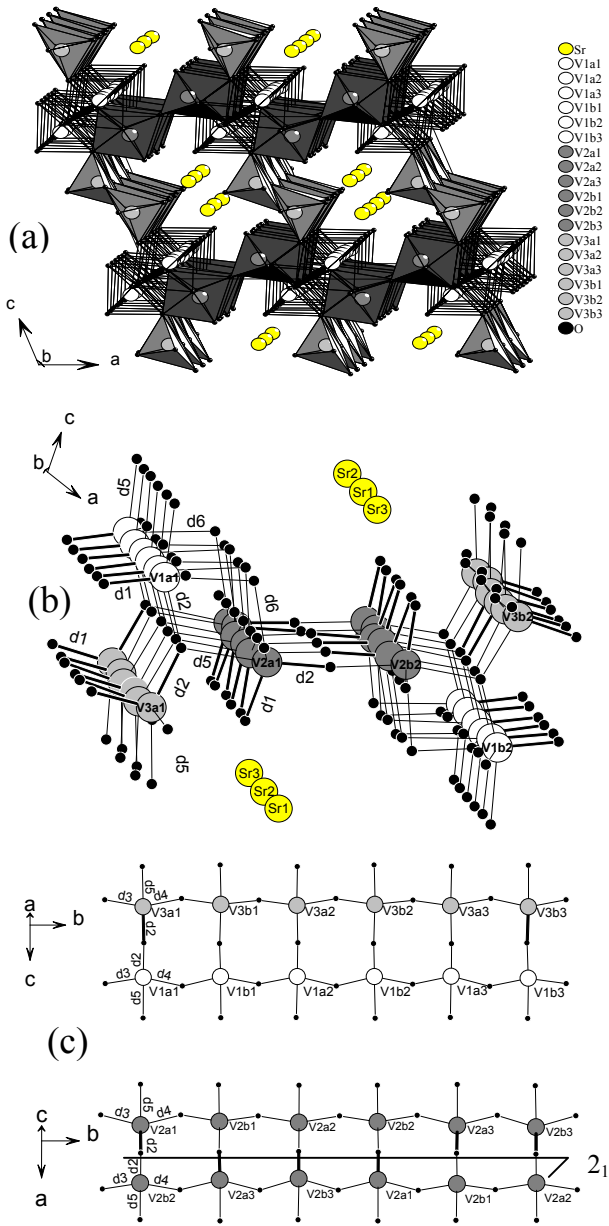

Figure 5: (a) Crystallographic structure of the LT SrV₆O₁₅ (T = 90K). (b) Crystallographic pattern of the low temperature structure. (c) Details of the crystallographic pattern. Short V-O bonds (< 1.77 Å) are indicated by thick lines.

## 4. Discussion

The *b*-axis threefold increase in a compound where the electronic filling is one 3*d* electron for three vanadium (average formal valence V⁴·⁶⁷⁺) strongly suggests that the structural transition is closely related with the charge distribution. The determination of the latter below the MIT is an important issue in SrV₆O₁₅. NMR results on SrV₆O₁₅ show only one signal above $T_{MIT}$, suggesting a rather uniform electronic state, while two signals, assigned to non-magnetic V⁵⁺ and magnetic V⁴⁺ sites respectively, can be observed below $T_{MIT}$ [7]. The Bond Valence Sum (BVS) analysis confirms that, unlike in the RT phase, a non-uniform charge state is achieved (see Table IV). It is however unclear from this analysis whether the charge differentiation is complete, resulting or not in a static distribution of V⁴⁺ and V⁵⁺. While the average BVS value for the vanadium is not far from the expected value (4.59 against 4.67), the BVS value for the different vanadium range only from 4.24 to 4.79, and not from 4 to 5. The hypothesis of a full charge differentiation has to be considered with caution, since an example of partial charge differentiation exists in the charge ordered (CO) state of the parent compound α'-NaV$_2$O$_5$ [17,18]. If one supposes however that a $V^{4+}$-$V^{5+}$ charge order with a full charge differentiation is established in the LT phase, the BVS results may provide interesting trends. The electronic filling of SrV$_6$O$_{15}$ (one $d$ electron for three vanadium) implies that the eighteen inequivalent vanadium are divided into twelve $V^{5+}$ and six $V^{4+}$ sites. The simple selection of the six vanadium with the lowest BVS values may not be the most reliable way to obtain the charge order due to the inaccuracy of the BVS method. A $V^{4+}$ state could however be unambiguously ascribe to three of the eighteen vanadium (V2a2, V3b1 and V3b2) which display a significantly lower BVS value that the others. Also it was shown in reference [16] from an analysis of more than one hundred known vanadium oxides that all the vanadium in trigonal bipyramid environment were in $V^{5+}$ valence state. It is obvious from Table IV that the five TB vanadium (V2a1, V2a3, V2b3, V3a1 and V3b3) have indeed the highest BVS values. If one ascribes a $V^{5+}$ state to the remaining vanadium with the BVS values above 4.66, one has to select three $V^{4+}$ among six vanadium. Magnetic properties of SrV$_6$O$_{15}$ moreover indicate the existence of a singlet ground state [7,8,19] that precludes the presence of isolated spin s = 1/2 $V^{4+}$. As a consequence, one is left with only four possible charge order patterns, shown in Fig. 7. A common trend of these possible charge orders is that the charge are mainly located on the V3 chains, unlike in NaV$_6$O$_{15}$ where a charge transfer from V3 chains to V1 or V2 sublattices occurs at the transition according to ref. [10]. Also, isolated clusters of (2 + 4) or 6 $V^{4+}$ are formed in SrV$_6$O$_{15}$ instead of infinite chains or ladders as previously proposed [8,19]. The presence of clusters of s = 1/2 $V^{4+}$ could be a feature common to all the $A$V$_6$O$_{15}$ in the charge order state, since the presence of clusters of three $V^{4+}$ was recently proposed in NaV$_6$O$_{15}$ [10]. However, the number of $V^{4+}$ within a cluster is twice as large in divalent $A$ cations compounds due to electronic filling reasons. Consequently, the even (odd) number of spin s = 1/2 within a cluster could explain the singlet (AFLRO) ground state of $A$V$_6$O$_{15}$ compounds with divalent (monovalent) $A$ cations. The previous analysis rests on the Bond Valence method, which absolute precision is rather poor. The five hypothesis of CO proposed must therefore be considered only as a guide for more direct determinations. Nevertheless the low BVS values of V2a2, V3b1 and V3b2 implies that they can be reliably considered as $V^{4+}$.

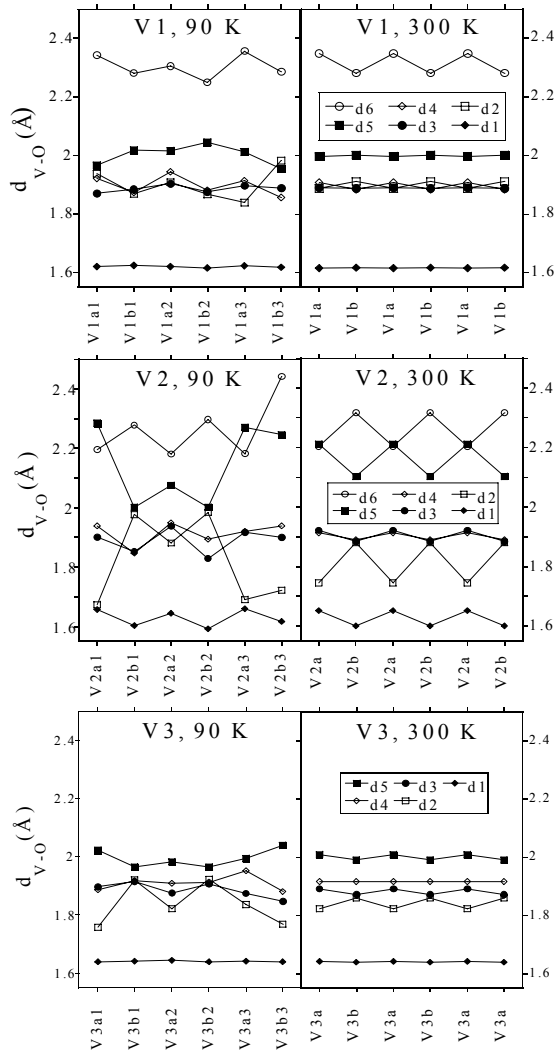

Figure 6: Left part: schematic representation of the V-O distances as a function of the vanadium site for the LT phase. Right part: evolution of the same distances above the transition. See Fig. 5b - 5c to locate the corresponding V-O bond in the structure.

Table IV: Bond valence Sum (BVS) on the vanadium of the room and low temperature structures. The equation used is: $s = \Sigma exp[-(R-R_0)/B]$ with values $R_0 = 1.790$, $B = 0.319$ [9]. The six oxygens in the first coordination sphere have been included in the calculation.

| | | | | | |
|---|---|---|---|---|---|
| 300 K | V1a | 4.58 | V2a | 4.59 | V3a | 4.51 |
| | V1b | 4.62 | V2b | 4.62 | V3b | 4.47 |
| 90 K | V1a1 | 4.54 | V2a1 | 4.79 | V3a1 | 4.79 |
| | V1b1 | 4.69 | V2b1 | 4.75 | V3b1 | 4.24 |
| | V1a2 | 4.41 | V2a2 | 4.28 | V3a2 | 4.57 |
| | V1b2 | 4.74 | V2b2 | 4.74 | V3b2 | 4.28 |
| | V1a3 | 4.62 | V2a3 | 4.73 | V3a3 | 4.46 |
| | V1b3 | 4.63 | V2b3 | 4.67 | V3b3 | 4.78 |

The origin of the MIT in mixed valence compounds is a long-standing but not fully resolved problem. The most famous example is the Verwey transition [20,21] in the magnetite Fe$_3$O$_4$. According to recent structural re-investigations [22], a charge ordering with a partial charge differentiation (Fe$^{2.5+} \rightarrow$ Fe$^{(2.5 \pm \delta)+}$, $\delta \approx 0.1$) occurs below the transition, without fulfilling the Anderson condition of minimum electrostatic repulsion [23]. An electronic instability opening a gap through a charge density wave (CDW) mechanism has been proposed, but the origin of the transition remains controversial in Fe$_3$O$_4$. The situation of β-SrV$_6$O$_{15}$ is somewhat analogous, since the proposed scenario of clusters of $V^{4+}$ clearly doesn't minimize the electrostatic repulsions. The 1D character of the electronic properties of the $A$V$_6$O$_{15}$ [3] may also favor a CDW mechanism [10]. Several studies, including band structure calculations, are under way to clarify this point.

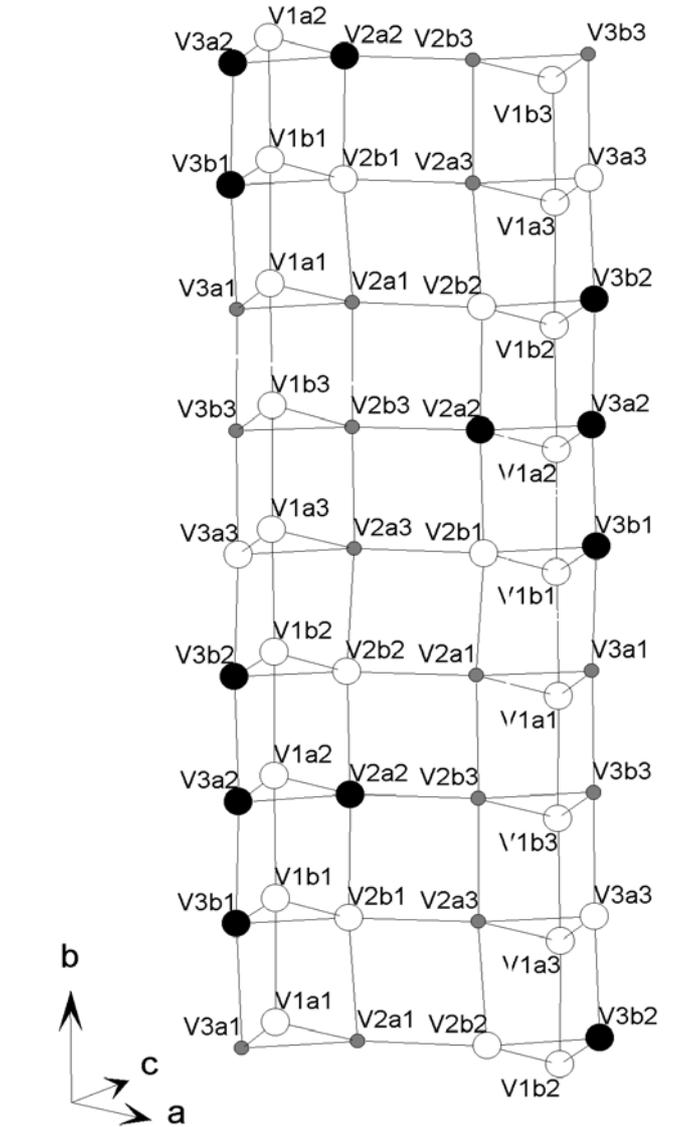

Fig. 7: charge order in SrV$_6$O$_{15}$ at low temperature shown on the crystallographic pattern introduced in Fig. 5b. The four $V^{4+}$ common to the four possible charge order are shown (dark balls). The four charge order patterns include this four $V^{4+}$ plus one of this pair: (V1a1, V1b3), (V1a3, V1b3), (V1a3, V3a3) or (V1a2, V3a3). The small grey balls represent the vanadium atoms in a trigonal bipyramid environment (see text).

## 5. Concluding remarks

In summary, we confirm the $2b$-superstructure in the space group $P2_1/a$ of $SrV_6O_{15}$ at 300 K. We established that a structural transition was associated with the metal-insulator of $SrV_6O_{15}$ at 170 K, with a $6b$-superstructure for the low temperature phase. A tendency of the low-T structure to charge separation and ordering in $V^{4+}$ clusters regularly spaced along the 1D $b$-axis is observed through a Bond Valence Sum analysis.

## Acknowledgements

We thank M. Evain for the selection of the single crystal, M. B. Lepetit for valuable discussions on charge ordering and C. Payen for fruitful discussions and the critical reading of manuscript.

## References

[1] Wadsley A.D., Acta Cryst. 8 (1955) 695.

[2] Bouloux J.C., Galy J., Hagenmuller P., Revue de Chimie Minérale 11 (1974) 48.

[3] Yamada H., Ueda Y., J. Phys. Soc. Japan 68 (1999) 2735-2740.

[4] Yamada H., Ueda Y., Physica B 284-288 (2000) 1651-1652.

[5] Ueda Y., Isobe M., Yamauchi T., J. Phys. Chem. Solids 63 (2002) 951-955.

[6] Yamauchi T., Ueda Y., Môri N., Phys. Rev. Lett. 89 (2002).

[7] Ueda Y., Yamada H., Isobe M., Yamauchi T., J. Alloys Compounds 317-318 (2001) 109.

[8] Isobe M., Ueda Y., Mol. Cryst. Liq. Cryst. 341 (2000) 271.

[9] Brown I.D., Structure and Bonding in Crystal, vol. 2, Academic Press, 1981, p. 2-31.

[10] Yamaura J.I., Isobe M., Yamada H., Yamauchi T., Ueda Y., J. Phys Chem. Solids 63 (2002) 957-960.

[11] Mentré O., Huvé M., Abraham F., J. Solid State Chem. 145 (1999) 186.

[12] Nonius BV, Delft, The Netherlands.

[13] Otwinowski Z, Minor W., Methods in Enzymology, vol. 276, Macromolecular Crystallography, Part A, edited by C. W. Carter Jr and R. M. Sweet, p. 307-326. New York: Academic Press.

[14] Petricek V. and Dusek M., JANA2000, Institute of Physics, Academy of Sciences of the Czech Republic, Prague, Czech Republic, 1998.

[15] X-SHAPE, Crystal Optimisation for Numerical Absorption Correction, STOE & Cie, Darmstadt (1996).

[16] Schindler M., Hawthorne F.C., Baur W.H., Chem. Mater. 12 (2000) 1248-1259.

[17] Fagot-Revurat Y., Mehring M., Kremer R.K., Phys. Rev. Lett. 84 (2000) 4176.

[18] Suaud N., Lepetit M.B., Phys. Rev. Lett. 88 (2002) 564051.

[19] Ueda Y., J. Phys. Soc. Jpn (Suppl. B) 69 (2000) 149-154.

[20] Verwey E. J. W., Nature 144 (1939) 327.

[21] Verwey E. J. W., Haayman P. W., Romeijan F. C., J. Chem. Phys. 15 (1947) 181.

[22] Wright J. P., Attfield J. P., Radaelli P. G., Phys. Rev. Lett. 87 (2001) 266401.

[23] Anderson P. W., Phys. Rev. 102 (1956) 1008.